\documentclass[12pt]{book}

\usepackage[dvips]{graphicx,color}
\usepackage{makeidx,tsukuba}

\makeauthorindex
\makeindex

\begin{document}

\BookTitle{\itshape The 28th International Cosmic Ray Conference}
\CopyRight{\copyright 2003 by Universal Academy Press, Inc.}
\pagenumbering{arabic}

\chapter{
Shock Acceleration and gamma radiation in Clusters of Galaxies}

\author{Pasquale Blasi,$^1$ and Stefano Gabici$^2$ \\
{\it (1) INAF, Osservatorio Astrofisico di Arcetri, Firenze, Italy\\
(2) Dipartimento di Astronomia, Universit\`a di Firenze, Firenze, Italy}
}

\section*{Abstract}

The nonthermal radiation observed from a handfull of clusters
of Galaxies is the proof that particle acceleration occurs in
the intracluster medium. It is often believed that shock surfaces
associated with either mergers of clusters of galaxies, or with 
the cosmological inflow of matter onto clusters during structure 
formation may be the sites for acceleration. We discuss here the
effectiveness of shock acceleration in the intracluster medium,
stressing that merger related shocks are typically weak, at least
for the so-called major mergers. We investigate the implications 
of shock strengths for gamma ray emission from single clusters 
of galaxies and for their detectability with future gamma ray experiments
such as GLAST. We also discuss the contribution of clusters of
galaxies to the extragalactic diffuse gamma ray background. 

\section{Merger shocks and particle acceleration}

Nonthermal radiation is observed from several clusters of galaxies,
showing that accelerated particles are present in the intracluster 
medium. Although the processes responsible for the acceleration are 
poorly known, there are some hints that they may be related to the
merging processes that build up the clusters from smaller substructures.
During cluster mergers, shock waves are formed due to the supersonic 
relative motion, and particle acceleration may take place through the first
order Fermi process [1,2]. In [3], shock acceleration during cluster 
mergers was investigated without {\it ad hoc} assumptions on the 
strength of the shocks. The compression factors of these shocks were 
instead calculated from the physical properties of clusters involved 
in the process of hierarchical structure formation.
In Fig. 1 we plot the distribution of Mach numbers of shocks formed during
mergers that end up in a cluster of mass $10^{15} M_\odot$ at the 
present cosmic time. This plot shows that the large majority of mergers 
have Mach numbers below 2, which correspond to spectra of accelerated 
particles which are much steeper that those that can explain nonthermal 
observations in the radio band. Only $\sim 6\%$ of the mergers have spectra 
flatter than $E^{-2.4}$ (corresponding to Mach numbers larger than 3), 
which suggests that Fermi acceleration at merger shocks may not be the 
main process for the generation of nonthermal particles in the intracluster
medium.
\begin{figure}[t]
  \begin{center}
    \includegraphics[height=13.5pc]{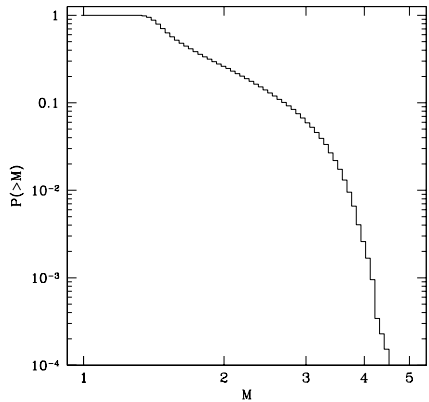}
  \end{center}
  \vspace{-0.5pc}
  \label{fig:mach}
  \caption{Distribution of Mach numbers during mergers of two clusters 
of galaxies, as obtained in [3].}
\end{figure}

\section{Clusters as single gamma ray sources}

Shock fronts are formed in clusters of galaxies either due to merger
events, or due to accretion of matter onto the potential well of a 
forming cluster. The latter shocks are called accretion shocks, and
by definition are very strong, since they propagate in a cold
non-virialized medium. Their Mach number can be up to several hundreds
and Fermi acceleration generates particle spectra $\propto E^{-2}$.
We concentrate here upon electrons accelerated at both merger and 
accretion shocks, and calculate the gamma ray emission that may be 
produced by these electrons as a result of their inverse Compton scattering
(ICS) on the cosmic microwave background (CMB) radiation. This mechanism 
is efficient if diffusion of the particles around the shock front is 
well described by B\"{o}hm diffusion and if the magnetic field is large 
enough in the shock upstream proximity. If these conditions are not 
fulfilled, it may happen that the maximum energy of the accelerated 
electrons is too low, and the CMB photons cannot be upscattered to 
gamma ray energies. In Fig. 2a we show the results of [4], where the 
$\log N-\log S$ for the gamma ray emission above 100 MeV 
from merging (dashed line) and accreting (solid line) clusters was 
calculated, following the merger and accretion histories.
\begin{figure}[t]
  \begin{center}
    \includegraphics[height=16.pc]{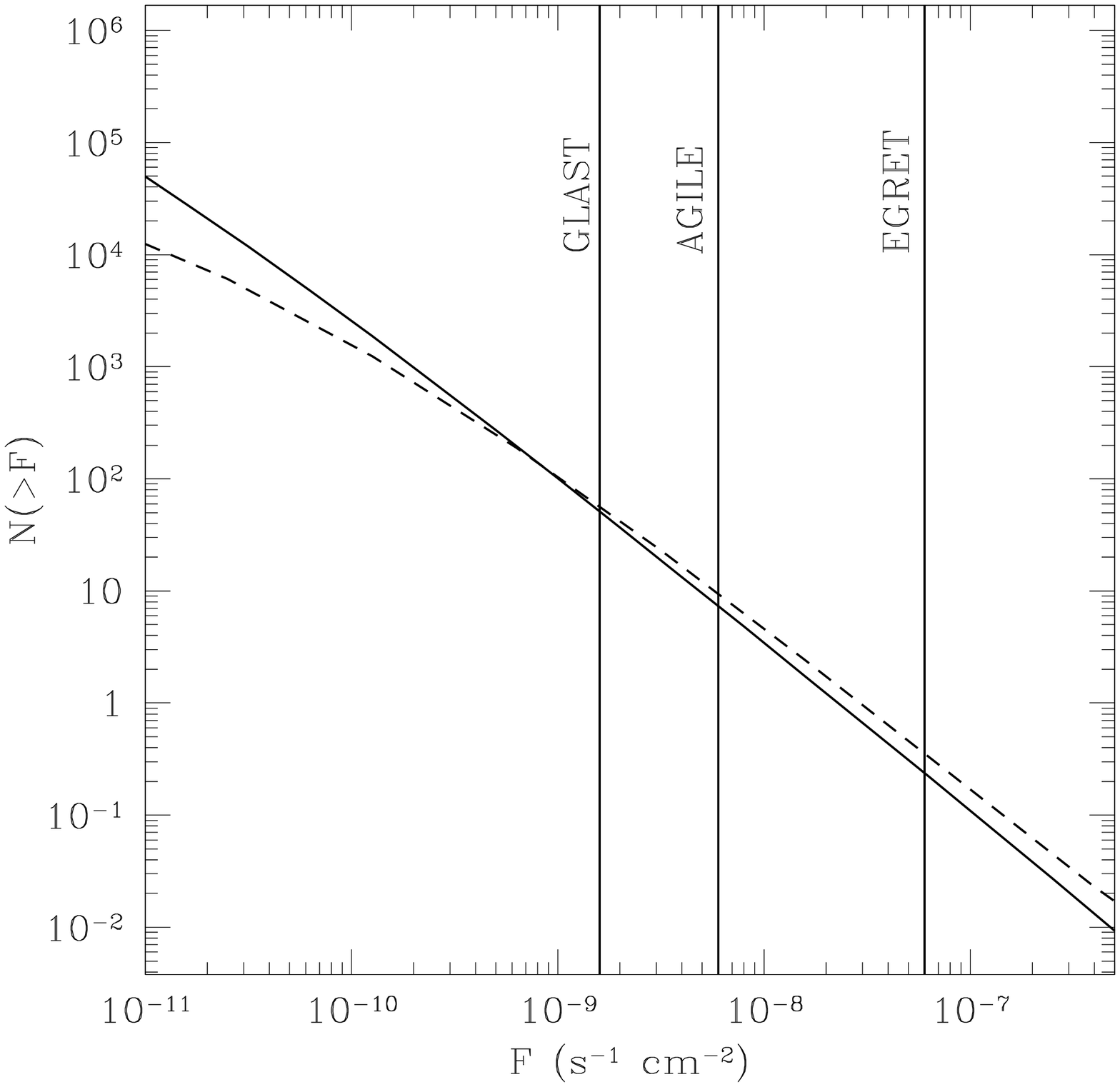}
    \includegraphics[height=16.pc]{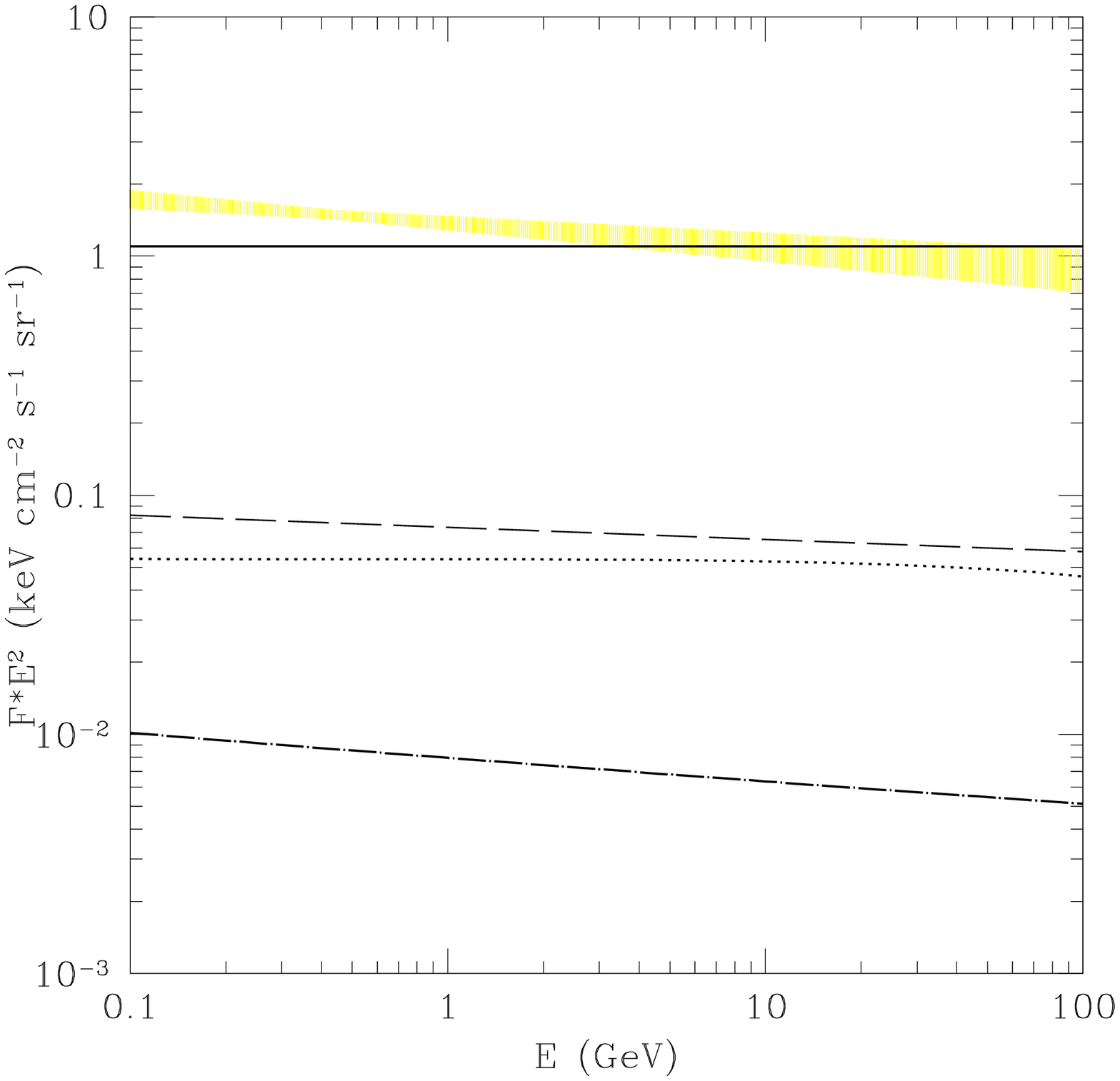} 
 \end{center}
  \vspace{-0.5pc}
  \label{fig:logNlogS}
  \caption{{\it a)} $Log N-Log S$ for mergers (dashed line) and for accretion 
(solid line). {\it b)} The diffuse gamma ray background from clusters of 
galaxies. See text for more information.}
\end{figure}
In the same plot we also indicated the sensitivity levels of gamma ray
telescopes such as EGRET, AGILE and GLAST. While no cluster was predicted
to be detectable by EGRET (and none has in fact been detected [5]), 
a few clusters should be visible with AGILE and $\sim 50$ clusters should be
identified by GLAST. The mergers are on average brighter, because they
release a large amount of energy in a short time (about a billion years),
but have spectra which are typically steep. Only for minor mergers, 
namely mergers involving two clusters with very different masses, the
spectra are flat enough to imply a considerable gamma ray emission. 
For particles accelerated at accretion shocks, the opposite is true:
accretion occurs over the all lifetime of a cluster and the energy 
involved is smaller than that released during a merger event. However,
the spectra of accelerated particles are as flat as ($E^{-2}$) and the 
gamma ray signal may be detectable. These
effects play together to make the $\log N-\log S$ curves for mergers
and accretion roughly similar in the region accessible to current
and future gamma ray telescopes. It is worth stressing however, that 
while mergers should become bright gamma ray sources only during the
merger event itself, accretion provides a steady gamma ray emission,
although limited to nearby clusters. Here we are neglecting the contribution
of hadronic gamma rays, namely those produced through production and
decay of neutral pions, that are the result of inelastic proton-proton
collisions. This contribution can be evaluated only by carefully accounting
for the injection of cosmic rays during the whole lifetime of the 
cluster, due to the confinement effect found in [6,7].

\section{The Diffuse Gamma Ray Background}

The superposition of the gamma ray emission from single clusters results in
a diffuse gamma ray emission. Our results [8] are plotted in 
Fig. 2b, where the dashed and dot-dashed lines represent 
the contribution of cluster mergers to the diffuse gamma ray background 
(DGRB) when the minimum mass of the merging clusters is taken to be 
$10^{12}$ and $10^{13} M_\odot$ and the shaded region reflects EGRET
observations [9]. Being $10^{12}M_\odot $ the mass of a galaxy, 
the shocks described here are not expected. In fact gas stripping is likely 
to occur in this case. We consider therefore the dot-dashed line as the 
most realistic prediction for the contribution of cluster mergers to the 
DGRB. The dotted line represents instead the contribution of accretion 
to the DGRB, and it is clear that this is the dominant contribution.
The solid line in Fig. 2b is the result of a previous
estimate [10] in which the shocks were all assumed strong (namely
with Mach numbers $\gg 1$). As we stressed above, this is not the case.
This, together with several other factors discussed in [8] explains 
their larger predicted DGRB. In all these curves the fraction 
of the kinetic pressure converted at the shocks into nonthermal electrons 
is taken to be $\sim 5\%$. 

\section{Conclusions}

In [3] we first found that merger related shocks were typically weak 
and the corresponding spectra of shock accelerated particles steep. We
included the effects of the stronger accretion shocks in [8].
This result that seemed initially in disagreement with the results of
N-body numerical simulations, that predicted a peak in the Mach number
distribution of merger shocks at $\sim 4$, was instead recently confirmed 
by improved numerical calculations [11].

Electrons accelerated at both merger and accretion shocks may upscatter 
the photons of the CMB to gamma ray energies and be detected by upcoming 
space-borne gamma ray telescopes, such as AGILE and GLAST. The diffuse 
gamma ray background generated by clusters of galaxies is about 10 times 
smaller that that observed by EGRET.

\section{References}
\vspace{\baselineskip}
\re
1.\ Fujita, Y., Sarazin, C. L. 2001, ApJ 563, 660
\re
2.\ Blasi, P. 2001, Astropart. Phys., 15, 223 
\re
3.\ Gabici, S., Blasi, P., 2003, ApJ 583, 695
\re
4.\ Gabici, S., Blasi, P., 2003 in preparation
\re
5.\ Reimer, O., Pohl, M., Sreekumar, P., Mattox, J.R., 2003, ApJ 588, 155
\re 
6.\ Berezinsky, V., Blasi, P., Ptuskin, V., 1997, ApJ 487, 529
\re
7.\ Volk, H. J., Aharonian, F. A., Breitschwerdt, D. 1996, Space Sci. Rev., 
75, 279
\re
8.\ Gabici, S., Blasi, P, 2003, Astropart. Phys., in press
\re
9.\ Sreekumar, P., et al., 1998, ApJ 494, 523 
\re
10.\ Loeb, A., Waxman, E., 2000, Nature 405, 156
\re
11.\ Ryu, D., Kang, H., Hallman, E., Jones, T.W., astro-ph/0305164

\endofpaper
\end{document}